\documentclass[11pt]{article}
\usepackage{Blois,epsfig}

\newcommand{\PL}[1]{{ Phys.\ Lett.\ } {\bf  #1}}

\newcommand{\PRL}[1]{{ Phys.\ Rev.\ Lett.\ } {\bf  #1}}

\def\be{\begin{equation}}
\def\ee{\end{equation}}
\def\bea{\begin{eqnarray}}
\def\eea{\end{eqnarray}}

\newcommand{\lsim}{\raise.3ex\hbox{$<$\kern-.75em\lower1ex\hbox{$\sim$}}}
\newcommand{\ima}{{\mbox{Im}\,}}
\newcommand{\rea}{{\mbox{Re}\,}}
\begin{document}
\vspace*{2cm}

\vspace*{2cm}
\title{COMMENTS ON THE LARGE $Nc$ BEHAVIOR OF LIGHT SCALARS}

\author{J.R. Pel\'aez}

\address{Departamento de F\'{\i}sica Te\'orica II. Universidad Complutense. 28040 Madrid .Spain.}

\maketitle\abstracts{
I review the large $N_c$ behavior  
of light resonances
generated from unitarized one-loop Chiral Perturbation Theory.
In contrast with the $\rho$ or $K^*$, the scalar behavior 
is at odds with a $\bar{q}q$ {\it dominant component}. In fact,
in the light scalar region, meson-meson amplitudes vanish as
$N_c$ increases. Also, the scalar widths, obtained from 
their associated poles, behave as $O(N_c^{1/2})<\Gamma<O(N_c)$.
We also clarify on the physical relevance of considering large $N_c$
not too far from real life, $N_c=3$, and the interpretation of the 
mathematical $N_c\rightarrow\infty$ limit.
}
\section{Introduction}
On the one hand, the large $N_c$ expansion \cite{'tHooft:1973jz}
is the only
analytic approximation to QCD in the whole
energy region, 
providing a clear definition of $\bar qq$ states, that become bound,
and whose
masses and widths behave as $O(1)$ and $O(1/N_c)$, respectively.
On the other hand, Chiral Perturbation Theory (ChPT) 
is the QCD low energy Effective Lagrangian
built as the most general derivative expansion
respecting SU(3) symmetry and 
containing only $\pi, K$ and $\eta$ mesons\cite{chpt1}. These particles 
are the QCD low energy degrees of freedom
since they are Goldstone bosons of the QCD spontaneous
chiral symmetry breaking.
For meson-meson scattering  ChPT is an expansion in even
powers of momenta, $O(p^2), O(p^4)$...,
 over a scale $\Lambda_\chi\sim4\pi f_0\simeq 1\,$GeV.
Since the $u$, $d$ and $s$ quark
masses are so small compared with $\Lambda_\chi$ they
are introduced as perturbations, giving rise to the 
$\pi, K$ and $\eta$ masses, counted as $O(p^2)$. 
At each order, ChPT is the sum of {\it all terms}
 compatible with the symmetries,
multiplied by ``chiral  parameters'', that absorb
loop divergences order by order, yielding finite results.
The leading order is universal, containing only one parameter,  $f_0$, 
that sets the scale of spontaneous symmetry breaking.
Different underlying dynamics manifest with different 
higher order  parameters, called $L_i$, that, once renormalized,
depend on a regularization scale as 
$
 L_i(\mu_2)=L_i(\mu_1)+\Gamma_i\log(\mu_1/\mu_2)/{16\pi^2},
$
where $\Gamma_i$ are constants\cite{chpt1}.
In physical observables the $\mu$ dependence is canceled
with that of the loop integrals.

The $\pi,K,\eta$ masses scale as $O(1)$ and 
$f_0$ as $O(\sqrt{N_c})$.
The $L_i$ parameters that determine
meson-meson scattering up to $O(p^4)$
 scale\cite{chpt1,chptlargen} as $O(N_c)$ for $i=1,2,3,5,8$
whereas
$2L_1-L_2, L_4,L_6$ and $L_7$ scale as $O(1)$.
In order to apply the large $N_c$ expansion, the $\mu$ scale, a dependence suppressed by $1/N_c$,
has to be  chosen\cite{chpt1} between $\mu=$0.5 and 1 GeV.
(In Figure 3 we will see that this estimate 
yields the correct behavior for light vector mesons,
firmly established as $\bar qq$ states).

In recent
years ChPT has been extended to higher energies by means of unitarization 
\cite{GomezNicola:2001as,Dobado:1996ps,Oller:1997ng,Pelaez:2003dy}. 
The main idea is that when projected into partial waves of definite
angular momentum $J$ and isospin $I$, physical amplitudes $t_{IJ}$ should satisfy
an elastic unitarity condition:
\begin{equation}
  \ima t_{IJ} =\sigma \vert t_{IJ}\vert^2 \quad\Rightarrow \quad\ima \frac{1}{t_{IJ}}=-\sigma \quad\Rightarrow\quad
t_{IJ}=\frac{1}{\rea t_{IJ}^{-1} - i \sigma}.
\end{equation}
Since the two body phase space  $\sigma$ is known, in order to have a unitary
amplitude we only need $\rea t^{-1}$, that can be obtained from ChPT: 
this is the
Inverse Amplitude Method (IAM) \cite{Dobado:1996ps,GomezNicola:2001as}.  
In this way, the IAM generates the $\rho$, $K^ *$, $\sigma$ and $\kappa$ resonances 
not initially present in ChPT, ensures unitarity
in the elastic region and respects the ChPT expansion.
When inelastic two-meson processes are present the IAM generalizes\cite{GomezNicola:2001as,Oller:1997ng} 
to $T\simeq(\rea T^{-1}-i \Sigma)^{-1}$, where 
$T$ is a matrix containing all partial waves 
between all physically accessible two-body states, whereas
$\Sigma$ is a diagonal
matrix with their phase spaces,
again well known. 
Using one-loop ChPT calculations,
the IAM provides  a remarkable description\cite{GomezNicola:2001as}
of  two-body $\pi$, K or $\eta$
scattering up to 1.2 GeV. In addition, it generates the $\rho$, $K^ *$, $\sigma$, $\kappa$,
$a_0(980)$,
$f_0(980)$ and the octet $\phi$. 
Such states are not included in the ChPT Lagrangian,
but each one has an associated pole
in the second Riemann sheet of its corresponding partial wave.
These poles appear already with the $L_i$ set used for standard ChPT,
and also with the $L_i$ set obtained from fits to data, which are compatible with each other.
For narrow, Breit-Wigner like, resonances, their mass and width
is roughly given by  $\sqrt{s_{pole}}\sim M_R-i\,\Gamma_R/2$.
Furthermore, the IAM respects the $O(p^4)$
correct low energy expansion, with chiral parameters
compatible with standard ChPT.
Different IAM fits\cite{GomezNicola:2001as} 
are mostly due to different ChPT truncation schemes,
equivalent up to $O(p^ 4)$.

Note that the ChPT amplitudes used are fully
renormalized, and therefore scale independent. 
There are no  cutoffs or a subtraction constants
where a spurious $N_c$ dependence could hide.
All the QCD $N_c$ dependence appears correctly
through the $L_i$, $f_0$ and the $\pi, K, \eta$ masses.

Recently\cite{Pelaez:2003dy}, 
by rescaling the ChPT parameters,  we have studied how
those generated resonances behave in the large $N_c$ expansion.
Thus, in Fig.1 we see what happens with the $\rho(770)$ and $K^*(892)$ vector mesons.
In real life, the modulus of their corresponding partial waves presents a peak,
that we have obtained from a fit to data,
that becomes narrower as $N_c$ increases whereas the mass remains almost the same.
By looking at the mass and width from the pole we see that, for both resonances,
they behave as expected for a $\bar{q}q$ state, i.e. $M\sim O(1)$, $\Gamma\sim O(1/N_c)$.

\hspace*{-.7cm}
\begin{minipage}{\textwidth}
\vspace{.3cm}
\begin{center}
  \psfig{figure=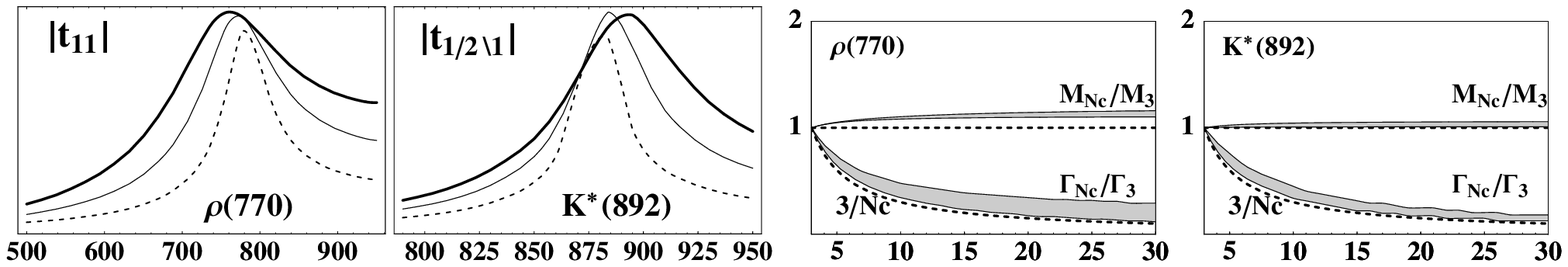,height=2.7cm}
\end{center} {\footnotesize Figure 1: Left:
    Modulus of $\pi\pi$ and $\pi K$ elastic amplitudes versus
    $\sqrt{s}$ for $(I,J)=(1,1),(1/2,1)$: $N_c=3$ (thick line),
    $N_c=5$ (thin line) and $N_c=10$ (dotted line), scaled at
    $\mu=770\,$MeV.  Right: $\rho(770)$ and $K^*(892)$ pole positions:
    $\sqrt{s_{pole}}\equiv M-i\Gamma/2$ versus $N_c$. The gray areas
    cover the uncertainty $N_c=0.5-1\,$GeV. The dotted lines show the
    expected $\bar q q$ large $N_c$ scaling.  }
\vspace{.3cm}
\end{minipage}

In contrast, in Figure 2 we see the behavior for the $\sigma$ (or $f_0(600)$) and the $\kappa$.
The results for the $a_0(980)$ and $f_0(980)$ are roughly similar, but 
more subtle\cite{Pelaez:2003dy}. 
It is evident that these {\it scalars behave completely different to $\bar{q}q$: The modulus of their
partial waves in the resonance region vanish and
their widths grow as $N_c$ increases}, as $O(N_c^{1/2})<\Gamma<O(N_c)$.

\hspace*{-.7cm}
\begin{minipage}{\textwidth}
\vspace{.3cm}
\begin{center}
\psfig{figure=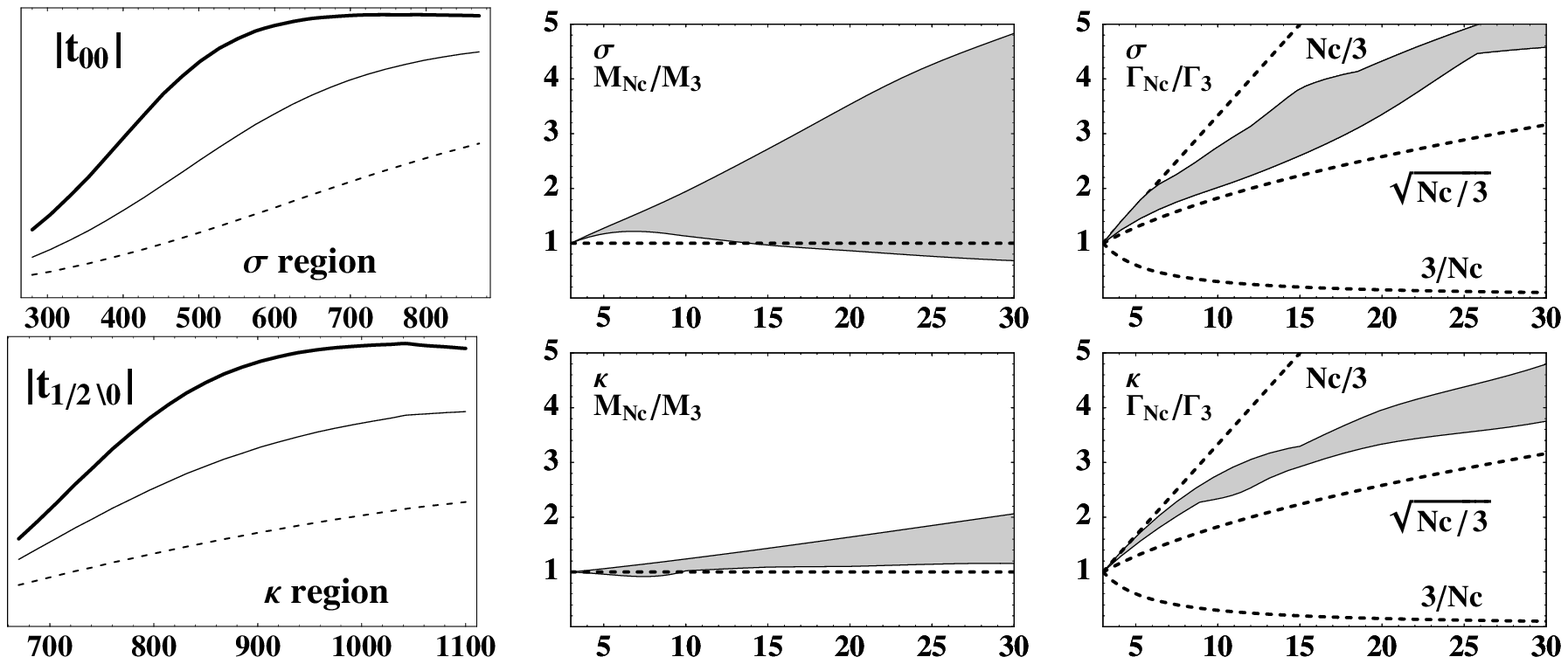,height=6cm}
\end{center}
{\footnotesize
Figure 2: (Top) Left:
Modulus of the $(I,J)=(0,0)$ scattering amplitude, versus $\sqrt{s}$
for $N_c=3$ (thick line), $N_c=5$ (thin line) and $N_c=10$ 
(dotted line), scaled at $\mu=770\,$MeV. Center: $N_c$ evolution
of the $\sigma$ mass. Right: $N_c$ evolution
of the $\sigma$ width. (Bottom): The same but for the $(1/2,0)$ amplitude
and the $\kappa$.
}
\vspace{.3cm}
\end{minipage}

\section{Discussion and conclusions}

We have seen that, within the unitarized ChPT approach,
 $\bar{q}q$ states are clearly identified 
whereas scalar mesons
behave differently as $N_c$ increases. Here I want to emphasize again
what can  and {\it what cannot} be concluded from this behavior
and clarify some frequent questions and misunderstandings that I have found in
private communications and the literature. 
\newcounter{input}
\begin{list}{$\bullet$}{
\setlength{\leftmargin}{0.2cm}
\setlength{\labelsep}{0.cm}}
\item {\it The \underline{dominant component}  of the $\sigma$ and $\kappa$ {\it in meson-meson scattering} does not behave as a $\bar{q}q$}.
\begin{list}{- }{
\setlength{\leftmargin}{0.2cm}
\setlength{\labelsep}{0.cm}}
  \item Why ``dominant''?
    Because, most likely, scalars are a mixture of different kind of
    states.  If the $\bar{q}q$ was {\it dominant}, they would behave
    as the $\rho$ or the $K^*$ in Figure 1.  {\it But it cannot be
      excluded that there is some smaller fraction of $\bar{q}q$.}
  \item Also, since scalars could 
be an admixture of states with different
nature and  wave functions, it could happen that
the small $\bar{q}q$ component could be concentrated in the core and better
seen in other reactions whereas 
in scattering we are seeing mostly the outer region.
\end{list}
  \item {\it Two meson and some tetraquark states\cite{Jaffe} have a consistent
      ``qualitative'' behavior}, i.e., both disappear in
    the continuum of the meson-meson scattering amplitude as $N_c$
    increases (also the glueballs for the $\sigma$ case but not for the $\kappa$).  Waiting for more quantitative results, we have not been able
     to establish yet the nature of that dominant component, but
      two-meson states or some kind of tetraquarks are, qualitatively,
      candidates to form that dominant component.
\end{list}

The IAM results
\cite{Pelaez:2003dy}, have been later confirmed, since
 ``very similar'' {\it numerical} results have been reported
 with other unitarization techniques \cite{Uehara:2003ax,Sun:2004de},
and the  the $N_c\rightarrow\infty$
limit has been studied\cite{Sun:2004de}.
This limit is interesting mathematically, and maybe could have some physical
relevance if the data and the large $N_c$ uncertainty on the
choice of scale were more accurate. Nevertheless
\begin{list}{\roman{input}.}{\usecounter{input}
\setlength{\leftmargin}{0.2cm}
\setlength{\labelsep}{0.cm}}
\item[$\bullet$ ] {\it Contrary to the large $N_c$ behavior \underline{in the vicinity of  $N_c=3$},
the mathematical $N_c\rightarrow\infty$ 
limit may not give information on the ``dominant component''
of light scalars.} The reason was 
commented above: In contrast to $\bar{q}q$
states, that become bound, 
 two-meson and some tetraquark
states dissolve in the 
continuum as $N_c\rightarrow\infty$. 
Thus, even if we started with an infinitesimal $\bar{q}q$ component
in a resonance, there could be a sufficiently large $N_c$ for which
it may become dominant, and beyond 
that $N_c$ the associated pole
would behave as a $\bar{q}q$ state
although the original state only had an infinitesimal admixture of $\bar{q}q$.
Also, since the mixings of 
different components could change
with $N_c$, a too large $N_c$ could alter significantly
the original mixings. 
\end{list}

Indeed this can happen\cite{Sun:2004de} for the $\sigma$ 
for {\it certain choices}
of chiral parameters: at a sufficiently high $N_c$ the pole 
may turn back toward
the real axis (see Figure 3). The IAM also yields such numerical result,
and for the $\kappa$ too. However, as commented above,
it does {\it not} mean that the ``correct interpretation... is that
the $\sigma$ pole is a conventional 
$\bar{q}q$ meson environed by heavy pion clouds''\cite{Sun:2004de}.
That the scalars are not conventional, is simply seen comparing
the scalars in Figure 2 with the ``conventional'' $\rho$ and $K^*$ in
Figure 1.
A large two-meson component is allowed, but
the $N_c\rightarrow\infty$  limit is not unique\cite{Sun:2004de} given the uncertainty
in the chiral parameters: scalar poles can move to negative 
mass square (quite weird), 
to infinity or to a  positive mass square. But even if 
the $L_i$ where determined with a much greater
precision, it is not clear that we could draw any conclusion
at $N_c\rightarrow\infty$:
As emphasized above and in\cite{Sun:2004de}, one loop ChPT
amplitudes are independent of the renormalization scale $\mu$, 
but the $L_i$ are not.
Thus, we have to choose a scale  between
roughly $0.5$ and 1 GeV to start our $N_c$ scaling. 
As seen in Figure 3, that uncertainty is enough
to change the $N_c\rightarrow\infty$ behavior, even when starting
from exactly the same set of $L_i$.

Therefore, robust 
conclusions on the dominant light scalar component,
can be obtained not too far from real life, $N_c=3$,  for a $\mu$ choice between
roughly $0.5$ and 1 GeV, and checking that simultaneously the $\rho$ and $K^*$ behave as almost pure $\bar{q}q$ states. That is one of the reasons why in
 Figures 1 and 2 we have only plotted up to 
$N_c=30$, but not 100, or a million.

\hspace*{-.7cm}
\begin{minipage}{\textwidth}
\vspace{.3cm}
\begin{center}
\psfig{figure=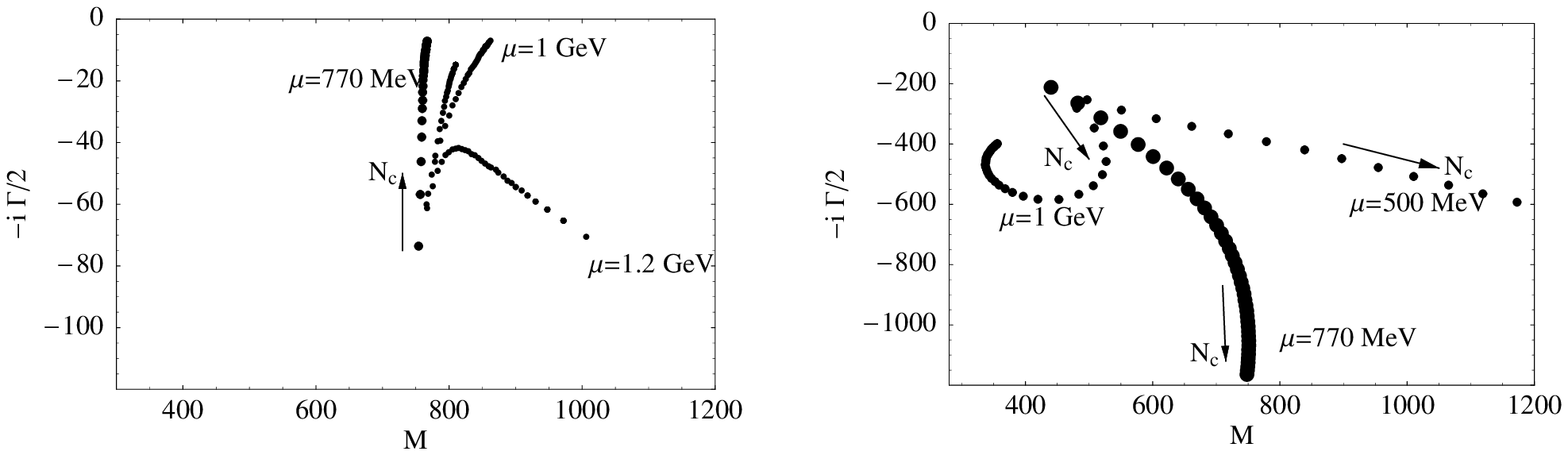,height=15cm,angle=-90}
\end{center}
{\footnotesize
Figure 3: Large $N_c$ behavior versus renormalization scale choice.
Left: The $\rho$ pole tends to the real 
axis if $0.5\hbox{GeV}<mu<1\hbox{GeV}$, but not for $\mu=1.2\,$GeV.
Right: The sigma pole behavior changes wildly for $\mu=1.2\,$GeV,
but always at odds with a $\bar{q}q$ dominant component. Note that the scale
here is larger than on the left.}
\vspace{.3cm}
\end{minipage}

{\it In summary, 
the dominant component of light scalars as generated from unitarized one loop ChPT scattering amplitudes does not behave as a
$\bar{q}q$ state as $N_c$ increases away from $N_c=3$.}

I thank the ``Rencontres de Blois'' organizers 
 for creating such a stimulating 
conference.

\end{document}